\documentclass[twocolumn,prl,showpacs,preprintnumbers,amsmath,amssymb]{revtex4-1}

\usepackage{graphicx}
\usepackage{dcolumn}
\usepackage{bm}

\begin{document}
\preprint{Phys. Rev. Lett. {\bf 113}, 069601 (2014)}

\title{Comment on ``Sticking of Hydrogen on Supported and Suspended Graphene at Low Temperature''}

\author{Dennis P. Clougherty}
\email{dpc@physics.uvm.edu}

\affiliation{
Department of Physics\\
University of Vermont\\
Burlington, VT 05405-0125}

\date{August 4, 2014}

\begin{abstract}
The sticking probability of cold atomic hydrogen on suspended graphene calculated by Lepetit and Jackson [Phys. Rev. Lett. {\bf 107}, 236102 (2011)] does not include the effect of fluctuations from low-frequency vibrations of graphene.  These fluctuations suppress the sticking probability for low incident energies  ($\lesssim 15$ meV).
\end{abstract}

\maketitle

Lepetit and Jackson (LJ) \cite{graphene} propose a model for the physisorption of atomic hydrogen on suspended graphene via phonon emission.  In their Fig.~2, a phonon density of states (DOS) $\rho(\omega)$ is plotted that vanishes linearly with $\omega$ at zero frequency.    LJ point out that in contrast to a constant DOS, a linear DOS eases a well-known divergence in the displacement autocorrelation function of the $n$th nearest neighbors $\langle (u_n-u_0)^2\rangle$ in two dimensions  \cite{mermin}.

In this Comment, the focus is on another   divergence.  The issue is already apparent in the Rayleigh-Schr\"odinger calculation of the shift in binding energy due to the atom-phonon interaction: a linear DOS, when combined with the frequency-dependent atom-phonon coupling $V_c$ in their Eq.~1, gives a (log) divergent correction to the binding energy at second-order.  

Neglecting contributions from the hydrogen continuum states, the second-order shift in the hydrogen binding energy $E^{(2)}_b$ from $V_c$ is given by $
E^{(2)}_b\approx \sum_{\bf Q} { |\langle b,0|V_c|b, {\bf Q}\rangle|^2/\hbar\omega_{\bf Q}}$
where $|b,0\rangle$ and $|b,{\bf Q}\rangle$ are eigenfunctions of the Hamiltonian \cite{jackson} $H_s+H_b$, the former is the ground state and the latter state has an excitation of wave vector $\bf Q$ in the bath.

In their Eq.~1, the summand of $V_c$ is inversely proportional to $\sqrt{\omega_{\bf Q}}$.  Hence in the continuum limit, $E^{(2)}_b\propto\int_0 d\omega\rho(\omega)/\omega^2$ which is log divergent for $\rho(\omega)\sim\omega$.  This divergence also appears in a perturbative calculation of the sticking probability at the two-loop level.

To keep the computation tractable, LJ truncates the phonon Fock space in two ways: (1) only zero and one-phonon states are included, and  (2) a low-frequency cutoff is used.  This gives an approximation to the one-loop atom self-energy $\Sigma(E)$.  With the omission of two-phonon states, $\Sigma(E)$ for the bound state propagator and vertex corrections are neglected in their numerical calculation.  

To estimate the effect of these omissions, $\Sigma(E)$ is calculated to the next order.  Consider the following Hamiltonian, closely related to that considered by LJ, that has been previously used to describe the sticking process \cite{dpc13}:
$H=H_p+H_b+H_c$
where
\begin{eqnarray}
H_p&=&E c_k^\dagger c_k -E_b c_b^\dagger c_b,\\
H_b&=&\sum_n{\hbar\omega_n {b_n^\dagger} b_n},\\
H_c&=&-(c_k^\dagger c_b+c_b^\dagger c_k)g_{kb}\sum_n \xi\ ({b_n+b_n^\dagger}) \nonumber\\
&&- c_b^\dagger c_b g_{bb}\sum_n \xi \ ({b_n+b_n^\dagger}) 
\label{gbb}
\end{eqnarray}
Here, $E_b$ is the atom binding energy in the static potential, and $E$ is the incident atom energy. The coupling parameters $\xi$, $g_{kb}$ and $g_{bb}$ and their numerical values are discussed in detail elsewhere \cite{dpc13}.  For the purposes of this Comment, it will be apparent that their values are not essential.

The atom amplitude in the entrance channel satisfies a Schr\"odinger-like equation with a generalized optical potential \cite{feshbach, dpc92} $\Sigma (E)$.

\begin{figure}
\includegraphics[width=8cm]{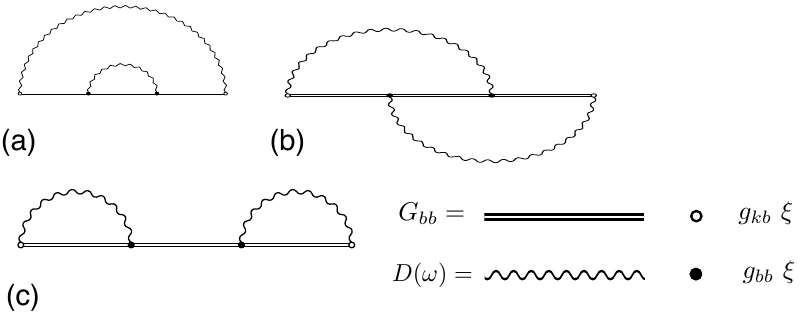}
\caption{\label{fig:se}  Diagrams contributing to the two-loop atom self-energy $\Sigma(E)$ to ${\cal O}(g_{kb}^2)$: (a) nested, (b) overlapping and (c) loop-after-loop.}
\end{figure}

$\Sigma (E)$ at one-loop is found to have finite real and imaginary parts; however, $\Sigma (E)$ at two-loops  ${\cal O}(g_{kb}^2)$ (analytically calculated from the three graphs in Fig.~\ref{fig:se}) is divergent.  The nested diagram for example gives the following contribution 
\begin{eqnarray}
\Sigma^{a}(E)&=&g_{kb}^2 g_{bb}^2 \xi^4\sum_{n, n'} {1\over (E-\omega_n+E_b+i\eta)^2}\nonumber\\
&&\times{1\over E-\omega_n-\omega_{n'}+E_b+i\eta}
\end{eqnarray}
The resulting integral is singular, with a divergence coming from processes involving a hard phonon ($\omega_n\sim E+E_b$) in the vicinity of the transition energy for adsorption and a soft phonon ($\omega_{n'}\sim 0$).  The overlap diagram is also divergent.
One must conclude that truncating the Feynman-Dyson perturbation expansion at the one-loop level is ill-advised for this model and non-perturbative approaches are needed.  

A continuum version of the LJ model for low-energy physisorption on a membrane under tension has recently been studied  \cite{dpc13}.  The analysis focuses on the effects of low-frequency fluctuations in the membrane on sticking.  In contrast to LJ's enhancement of sticking, I find {\it zero} probability of sticking for incident energies below 15 meV.  The effects from a finite sample size are negligible for $\mu$m-sized samples \cite{dpc13}.

Support by the National Science Foundation (No. DMR-1062966) is gratefully acknowledged.

\bibliography{qs}

\end{document}